\documentclass[twocolumn,prl,showpacs,amsfonts,amsmath,assymb,eufrak]{revtex4}
\usepackage{amsmath,amsfonts,amssymb,makeidx,ifthen}
\usepackage[dvipdfmx]{graphicx}
\usepackage[dvipdfmx]{hyperref}\hypersetup{hidelinks}

\makeatletter
\long\def\@makecaption#1#2{{\small
\advance\leftskip1cm
\advance\rightskip1cm
\vskip\abovecaptionskip
\sbox\@tempboxa{#1: #2}%
\ifdim \wd\@tempboxa >\hsize
 #1: #2\par
\else
\global \@minipagefalse
\hb@xt@\hsize{\hfil\box\@tempboxa\hfil}%
\fi
\vskip\belowcaptionskip}}
\makeatother
\def\eq#1\en{\begin{equation}#1\end{equation}}  
\def\eqa#1\ena{\begin{align}#1\end{align}}
\def\eqg#1\eng{\begin{gather}#1\end{gather}}
\newcommand{\lb}[1]{\label{e:#1}}
\newcommand{\rlb}[1]{\eqref{e:#1}} 
\newcommand{\nl}{\notag\\}

\newcommand{\sumtwo}[2]%
{\mathop{\sum_{#1}}_{#2}}
\newcommand{\sumthree}[3]%
{\mathop{\mathop{\sum_{#1}}_{#2}}_{#3}}
\newcommand{\sumfour}[4]%
{\mathop{\mathop{\mathop{\sum_{#1}}_{#2}}_{#3}}_{#4}} 
\newcommand{\prodtwo}[2]%
{\mathop{\prod_{#1}}_{#2}}
\newcommand{\mintwo}[2]%
{\mathop{\min_{#1}}_{#2}}
\newcommand{\maxtwo}[2]%
{\mathop{\max_{#1}}_{#2}}
\newcommand{\maxthree}[3]%
{\mathop{\mathop{\max_{#1}}_{#2}}_{#3}}
\newcommand{\limtwo}[2]%
{\mathop{\lim_{#1}}_{#2}}
\newcommand{\suptwo}[2]%
{\mathop{\sup_{#1}}_{#2}}
\newcommand{\supthree}[3]%
{\mathop{\mathop{\sup_{#1}}_{#2}}_{#3}}
\newcommand{\supfour}[4]%
{\mathop{\mathop{\mathop{\sup_{#1}}_{#2}}_{#3}}_{#4}} 
\newcommand{\inftwo}[2]%
{\mathop{\inf_{#1}}_{#2}}
\newcommand{\infthree}[3]%
{\mathop{\mathop{\inf_{#1}}_{#2}}_{#3}}
\newcommand{\inffour}[4]%
{\mathop{\mathop{\mathop{\inf_{#1}}_{#2}}_{#3}}_{#4}} 

\newcommand{\bss}{\boldsymbol{\sigma}}

\newcommand{\bbC}{\mathbb{C}}

\newcommand{\bbR}{\mathbb{R}}
\newcommand{\bbZ}{\mathbb{Z}}

\newcommand{\up}{\uparrow}

\newcommand{\qedm}{\rule{1.5mm}{3mm}}

\newcommand{\Sz}{\hat{S}^{{\rm z}}}
\newcommand{\hA}{\hat{A}}
\newcommand{\hH}{\hat{H}}
\newcommand{\hU}{\hat{U}}
\newcommand{\hV}{\hat{V}}
\newcommand{\hR}{\hat{R}}
\newcommand{\hh}{\hat{h}}
\newcommand{\DE}{\mathit{\Delta}E}
\newcommand{\ZZ}{\bbZ_2\times\bbZ_2}
\newcommand{\HH}{\hat{H}_{\rm H}}
\newcommand{\HAKLT}{\hat{H}_{\rm AKLT}}
\newcommand{\Htr}{\hat{H}_{\rm tr}}
\newcommand{\bS}{\hat{\boldsymbol{S}}}
\newcommand{\GSs}{\ket{\Phi^{\rm gs}_s}}

\newcommand{\bra}[1]{\langle#1|}
\newcommand{\ket}[1]{|#1\rangle}

\newcommand{\para}[1]{{\em #1}\/.---}
\newcommand{\midskip}{\vspace{3pt}}

\begin{document}
\title{Topological phase transition and $\bbZ_2$ index for $S=1$ quantum spin chains}

\author{Hal Tasaki}
\affiliation{
Department of Physics, Gakushuin University, 
Mejiro, Toshima-ku, Tokyo 171-8588, Japan}

\date{\today}

\begin{abstract}
We study $S=1$ quantum spin systems on the infinite chain with short ranged Hamiltonians which have certain  rotational and discrete symmetry.
We define a $\bbZ_2$ index for any gapped unique ground state, and prove that it is invariant under smooth deformation.
By using the index, we provide the first rigorous proof of the existence of a ``topological'' phase transition, which cannot be characterized by any conventional order parameters, between the AKLT model and trivial models.
This rigorously establishes that the AKLT model is in a nontrivial symmetry protected topological phase.
\end{abstract}


\maketitle

\para{Introduction and motivation}%
In early 1980's, Haldane discovered that the antiferromagnetic Heisenberg chain with the Hamiltonian $\HH=\sum_j\bS_j\cdot\bS_{j+1}$ has a unique gapped ground state when spin $S$ is an integer \cite{Haldane1981,Haldane1983a,Haldane1983b,Affleck1985,Affleck1989}.
The discovery opened a rich area of research in quantum many-body systems.
See, e.g., \cite{ZengChenZhouWenBOOK,TasakiBOOK}.
After the validity of Haldane's conclusion had been confirmed, an important issue was to elucidate the true nature of the gapped ground states of $\HH$.
A typical problem was to precisely characterize the difference between the gapped ground state of the solvable AKLT model  $\HAKLT=\sum_j\{\bS_j\cdot\bS_{j+1}+(\bS_j\cdot\bS_{j+1})^2/3\}$  \cite{AKLT1,AKLT2} and the trivial gapped ground state (where all spins are in the 0 state) of the trivial model $\Htr=\sum_j(\Sz)^2$, both for the $S=1$ chain.
The two ground states cannot be distinguished by any conventional order parameters.
It was soon realized that the ground states in the ``Haldane phase'', to which $\HAKLT$ and $\HH$ belong,  are ``exotic'' in the sense that they exhibit hidden antiferromagnetic order that can be characterized by a string order parameter \cite{AKLT2,denNijsRommelse}, and are accompanied by edge spins when defined on open chains \cite{AKLT2,Kennedy1990}.
These exotic properties, as well as the existence of a gap, were then interpreted as natural consequences of breakdown of hidden $\ZZ$ symmetry \cite{KennedyTasaki1992A,KennedyTasaki1992}.

It gradually became clear however that the picture of hidden $\ZZ$ symmetry breaking was neither sufficient nor necessary to characterize the Haldane phase \cite{Oshikawa92}.
In 2009, Gu and Wen pointed out that the Haldane phase should be identified as a symmetry protected topological phase \cite{GuWen2009}.
This means that, for example, 
$\HAKLT$ and $\Htr$ can be smoothly connected through a family of Hamiltonians with a gapped unique ground state if any short ranged Hamiltonian is available;
if, on the other hand, proper symmetry is imposed on the family of accessible Hamiltonians, then one must go through a phase transition in order to connect $\HAKLT$ and $\Htr$.
Such a phase transition is called ``topological'' since it is not characterized by a conventional order parameter.
A complete set of symmetry required to protect the Haldane phase was soon identified by Pollmann, Turner, Berg, and Oshikawa \cite{PollmannTurnerBergOshikawa2010,PollmannTurnerBergOshikawa2012}.
They concluded that the Haldane phase in odd $S$ quantum spin chains is protected either by (S1)~$\ZZ$ symmetry (i.e., the $\pi$ rotations about the x and z axes), (S2)~bond-centered reflection (inversion) symmetry, or (S3)~time-reversal symmetry.
They also showed that, for each symmetry, the Haldane phase and the trivial phase can be distinguished by a $\bbZ_2$ index which identifies the projective representation of the symmetry group, provided that the ground states are represented as (injective) matrix product states \cite{FannesNachtergaeleWerner1992}.

All these pictures suggest that the infinite-volume ground states of the one-parameter family of Hamiltonians 
\eq
\hH_s=s\HAKLT+(1-s)\Htr,
\lb{Ht}
\en
with $s\in[0,1]$, should exhibit a  ``topological" phase transition between the trivial phase with small $s$ and the Haldane phase with $s$ close to 1.
But, rather surprisingly,  the existence of a phase transition was not rigorously established before.
Although some of the known arguments are very plausible, there are still delicate gaps between mathematically rigorous proofs, as we now discuss briefly.
See \cite{TasakiNext} for details.
(On the other hand there are rigorous and explicit results which show that certain seemingly different ground states can be smoothly connected with each other.  See, e.g., \cite{PollmannTurnerBergOshikawa2012,BachmannNachtergaele2014}.)
\par\noindent
(i)~{\em Change in parity}\/: In a periodic chain with an odd number of sites, the ground states of $\HAKLT$ and $\Htr$ have odd and even parities, respectively, with respect to the reflection about a single bond \cite{PollmannTurnerBergOshikawa2012}.
Thus there must be a level crossing at intermediate $s$, suggesting a phase transition.
But this does not really imply a phase transition in the infinite volume limit.
The situation is even trickier since there is no level crossing in a chain with an even number of sites.
\par\noindent
(ii)~{\em Change in indices}\/:
The indices characterizing projective representations of the group symmetry provide a sophisticated support for  the existence of a phase transition \cite{PollmannTurnerBergOshikawa2010,PollmannTurnerBergOshikawa2012,ZengChenZhouWenBOOK,TasakiBOOK}.
There are three indices (corresponding to the three types of symmetry) which take $1$ and $-1$ in the ground states of $\Htr$ and $\HAKLT$, respectively.
Then there should be a phase transition associated with a jump in an index.
A major drawback of this approach is that the indices are well defined only for matrix product states (which also satisfies a strong condition called injectivity).
Although it is known that  a gapped ground state can be efficiently approximated by a matrix product state (see, e.g., \cite{AradLandauVaziraniVidick2017}) the approximation is not precise enough to derive a definite conclusion about phase transitions in full ground states.
The same comment applies to the non-local order parameters discussed in  \cite{Haegeman2012,PollmannTurner2012} (e.g., (20) of \cite{PollmannTurner2012}).
These quantities are well defined for a general state, but their quantization can be proved only for matrix product states.
\par\noindent
(iii)~{\em Hidden $\ZZ$ symmetry breaking}\/:
There is a nonlocal unitary transformation (for open chains) which maps the Hamiltonians  \rlb{Ht} to different local Hamiltonians \cite{KennedyTasaki1992A,KennedyTasaki1992,Oshikawa92}.
After the transformation the number of infinite volume ground states of the models with $s=0$ and 1 become one and four, respectively.
Then, by definition, the infinite volume limit of the transformed model must exhibit a phase transition.
This rigorous result strongly suggests that the original model too undergoes a phase transition.
But we still do not have any proof in this direction.
This is because the original model always has a unique ground state, and we do not know anything about the nature of the phase transition in the transformed model (except for the change in the number of ground states).

One of the contributions of the present work is the first completely rigorous proof of the existence of a phase transition in \rlb{Ht}.
More generally, we establish that the AKLT model is in a nontrivial symmetry protected topological phase within three classes, called C1, C2, and C3, of Hamiltonians that we specify below.
The result extends to other quantum spin chains and one-dimensional electron systems with proper symmetry \cite{TasakiNext}.
Our proof, which is based on a new ``topological'' index defined for a unique gapped ground state of an infinite chain, clearly illustrates why and how a gapless point emerges when the index changes.
We hope that this interesting argument leads to a deeper understanding of topological phase transitions.
Our index is related to the order parameter introduced by Nakamura and Todo \cite{NakamuraTodo2002}.
A different index was defined in \cite{BachmannNachtergaele2014A} also for infinite systems, but it has not yet been used to analyze phase transitions.

\midskip\para{Basic strategy}%
Let us briefly (and heuristically) discuss the basic idea of our proof for a finite system.
The proof is by contradiction.
Consider a large periodic chain, and let $\GSs$ be the ground state of $\hH_s$ of \rlb{Ht} for $s\in[0,1]$.
We assume that the ground state is always unique and accompanied by a nonvanishing gap.
We define the twist operator of Affleck and Lieb \cite{AffleckLieb1986},
\eq
\hV_\ell=\bigotimes_{j:|j-\frac{1}{2}|\le\ell+\frac{1}{2}}\exp\Bigl[-i\,2\pi\frac{j+\ell}{2\ell+1}\Sz_j\Bigr],
\lb{V0}
\en
which is the local version of the twist operator of Lieb, Schultz, and Mattis \cite{LiebSchultzMattis1961}.
Following Nakamura and Voit \cite{NakamuraVoit2002}, Nakamura and Todo \cite{NakamuraTodo2002} pointed out that the expectation value of the twist operator acts as an order parameter for the Haldane phase (see also \cite{Bonesteel}).
In the present context, the ground states at $s=0$ and 1 are characterized by $\bra{\Phi^{\rm gs}_0}\hV_\ell\ket{\Phi^{\rm gs}_0}=1$ and $\bra{\Phi^{\rm gs}_1}\hV_\ell\ket{\Phi^{\rm gs}_1}\simeq-1$ for sufficiently large $\ell$.
We also show from the symmetry that $\bra{\Phi^{\rm gs}_s}\hV_\ell\ket{\Phi^{\rm gs}_s}$ is always real.
Then, by continuity, there must be $s$ such that $\bra{\Phi^{\rm gs}_s}\hV_\ell\ket{\Phi^{\rm gs}_s}=0$.
This means that $\ket{\Psi}=\hV_\ell\ket{\Phi^{\rm gs}_s}$ is orthogonal to $\GSs$.
From the variational estimate of \cite{LiebSchultzMattis1961}, we also see that $\bra{\Psi}\hH\ket{\Psi}-E^{\rm gs}_s=O(\ell^{-1})$, where $E^{\rm gs}_s$ is the ground state energy of $\hH_s$.
We thus conclude that the energy gap of $\hH_s$ is $O(\ell^{-1})$.
But this is a contradiction since $\ell$ can be made as large as one wishes.
Note that, in this argument, the twist operator $\hV_\ell$ plays two essentially different roles, one as an observable whose expectation value is an order parameter, and the other as an unitary operator which generates a low energy excited state exactly as in the original work of Lieb, Schultz, and Mattis \cite{LiebSchultzMattis1961}.

By using a similar idea we can also show for a unique gapped ground state $\ket{\Phi^{\rm gs}}$ (with a certain symmetry condition) that the expectation value $\bra{\Phi^{\rm gs}}\hV_\ell\ket{\Phi^{\rm gs}}$ takes a constant sign for sufficiently large $\ell$.
We identify the sign as a $\bbZ_2$ index of the ground state.
Our index is closely related to the Berry phase of quantum spin chains introduced by Hatsugai
\cite{en:Hatsugai,Hatsugai2006,Hatsugai2007,HiranoKatsuraHatsugai2008,HiranoKatsuraHatsugai2008B}, and also to the polarization in electron systems.
For the latter, see \cite{NakamuraVoit2002,WatanabeOshikawa2018} and references therein.

\midskip\para{Setting and results}%
We study $S=1$ quantum spin systems on the infinite chain $\bbZ$.
By $\hat{S}^{\alpha}_j$ with $\alpha=\rm x,y,z$ we denote the $\alpha$-component of the spin operator at site $j\in\bbZ$.
We denote by $\hU^\alpha_\theta=\bigotimes_{j=-\infty}^\infty e^{-i\theta\hat{S}^{\alpha}_j}$ the global rotation by $\theta\in\bbR$ about the $\alpha$-axis.

We define three classes, which we call C1, C2, and C3, of Hamiltonians.
A Hamiltonian in these classes is written as
$\hH=\sum_{j=-\infty}^\infty\hh_j$,
where the local Hamiltonian $\hh_j$ depends only on spin operators at sites $k$ such that $|j-k|\le r$, and satisfies $\Vert\hh_j\Vert\le h_0$ and $(\hU^{\rm z}_\theta)^\dagger\hh_j\hU^{\rm z}_\theta=\hh_j$ for any $\theta$.
The range $r$ and $h_0$ are arbitrary fixed positive constants.
The Hamiltonian is invariant under any rotation about the z-axis.
We  require additional discrete symmetry depending on the class.
In C1, we assume that the Hamiltonian is invariant under the $\pi$-rotation about the x-axis, i.e., $\hU^{\rm x}_\pi\hH\hU^{\rm x}_\pi=\hH$.
In C2, we assume reflection invariance $\hR\hH\hR=\hH$, where $\hR$ is the bond-centered reflection operator induced by $j\to 1-j$, i.e., $\hR\hat{S}^{\alpha}_j\hR=\hat{S}^{\alpha}_{1-j}$.
In C3, we assume that $\hH$ is invariant under time-reversal $\hat{S}^{\alpha}_j\to-\hat{S}^{\alpha}_j$.

Let us summarize standard definitions of the uniqueness of the ground state and of the energy gap for infinite systems \cite{AffleckLieb1986,AKLT2}.
Given a Hamiltonian $\hH$
for the infinite chain, consider a corresponding Hamiltonian $\hH_L=(\sum_{j=-(L-r)}^{L+1-r}\hh_j)+\mathit{\Delta}\hh_{-L}+\mathit{\Delta}\hh_{L+1}$ on a finite chain $\{-L,\ldots,L+1\}$, where $\mathit{\Delta}\hh_{-L}$, $\mathit{\Delta}\hh_{L+1}$ are certain boundary Hamiltonians which act on spins around $-L$ and $L+1$, and respect the symmetry in each class.
Let $\ket{\Phi_L^{\rm gs}}$ be a ground state of  $\hH_L$.
The (infinite volume) ground state $\omega(\cdot)$ of the Hamiltonian $\hH$  is defined as the limit
\eq
\omega(\hA)=\lim_{L\up\infty}\bra{\Phi_L^{\rm gs}}\hA\ket{\Phi_L^{\rm gs}},
\lb{omega}
\en
where $\hA$ is an arbitrary local operator.
(By a  local operator we mean a function of a finite number of spin operators.)
We say that $\hH$ has a unique (infinite volume) ground state if the limiting $\omega(\cdot)$ is independent of the choice of the boundary Hamiltonians $\mathit{\Delta}\hh_{\pm L}$ and the choice of the finite volume ground state $\ket{\Phi_L^{\rm gs}}$.

Suppose that $\hH$ has a unique ground state $\omega(\cdot)$.
We say that $\hH$ has a nonvanishing energy gap if there is a constant $\epsilon>0$, and one has
\eq
\omega(\hA^\dagger[\hH,\hA])\ge\epsilon,
\lb{gap}
\en
for any local operator $\hA$ such that $\omega(\hA)=0$ and $\omega(\hA^\dagger\hA)=1$.
The supremum of such $\epsilon$, which we denote as $\DE$, is the energy gap of $\hH$.
By recalling the definition \rlb{omega} of the ground state, one sees that this is nothing but a straightforward extension of the standard variational characterization of the energy gap.
Note also that, although $\hH$ acts on infinitely many spins, the commutator  $[\hH,\hA]$ is a local operator.

Suppose that $\omega(\cdot)$ is a unique ground state of a Hamiltonian in class C1.
The uniqueness implies that the state is invariant under  $\pi$-rotation about the x-axis, i.e., $\omega(\hU^{\rm x}_\pi\hA\hU^{\rm x}_\pi)=\omega(\hA)$ for any local operator $\hA$.
Since $\hU^{\rm x}_\pi\hV_\ell\hU^{\rm x}_\pi=\hV_\ell^\dagger$, we see that $\omega(\hV_\ell)\in\bbR$.
Similarly the unique ground state $\omega(\cdot)$ of $\hH$ in C2 satisfies $\omega(\hR\hA\hR)=\omega(\hA)$ for any $\hA$.
Since $\exp[i\,2\pi\Sz_j]=1$, we find
\eqa
\hR\hV_\ell\hR&=\bigotimes_{j}\exp\Bigl[-i\,2\pi\frac{1-j+\ell}{2\ell+1}\,\Sz_j\Bigr]
\nl&=\bigotimes_{j}\exp\Bigl[i\,2\pi\frac{j+\ell}{2\ell+1}\,\Sz_j\Bigr]=\hV_\ell^\dagger,
\lb{RVR}
\ena
which again implies $\omega(\hV_\ell)\in\bbR$.
We can also show  $\omega(\hV_\ell)\in\bbR$ for the class C3.  See below.

\para{Theorem 1}%
Suppose that a Hamiltonian $\hH$ in C1, C2 or C3 has a unique ground state $\omega(\cdot)$ and a gap $\DE>0$.
Then for any $\ell$ such that $\ell>\max\{\ell_0,C/\DE\}$, the expectation value $\omega(\hV_\ell)\in\bbR$ is nonzero and has a constant sign.
Here $C$ and $\ell_0$ are positive constants which depend only on the constants $r$ and $h_0$ (which we fixed in the beginning).

The theorem guarantees that we can unambiguously define an index $\sigma(\hH)=\pm1$ for a Hamiltonian $\hH$ with a unique gapped ground state by
\eq
\sigma(\hH)=\frac{\omega(\hV_\ell)}{\bigl|\omega(\hV_\ell)\bigr|}\quad \text{for}\ \ell>\max\Bigl\{\ell_0,\frac{C}{\DE}\Bigr\}\,.
\en
One can also prove that $\omega(\hV_\ell)\to\pm1$ as $\ell\up\infty$ by using the method in \cite{Tasaki2017A}.

To state the essential property of the index, which is Theorem~2, we introduce the (standard) notion that two Hamiltonians are smoothly connected.

\para{Definition}%
Two Hamiltonians $\hH_0$ and $\hH_1$ in one of the classes C1, C2, or C3 are said to be smoothly connected within the class when the following is valid.
There are a positive constant $\DE_{\rm min}$ and a one-parameter family of Hamiltonians $\hH_s$ (with $s\in[0,1]$) in the same class.
For each $s\in[0,1]$, the Hamiltonian $\hH_s$ has a unique infinite volume ground state $\omega_s(\cdot)$ with a nonvanishing energy gap which is not less than $\DE_{\rm min}$.
For any local operator $\hA$, the expectation value $\omega_s(\hA)$ is continuous in $s\in[0,1]$.

\para{Theorem 2}%
If two Hamiltonians $\hH_0$ and $\hH_1$  in one of the classes C1, C2, or C3 are smoothly connected within the class, then $\sigma(\hH_0)=\sigma(\hH_1)$.

Thus our index is ``topological'' in the sense that it is invariant under  smooth deformation.
A trivial but important corollary is the following.

\para{Corollary 1}%
If one has $\sigma(\hH_0)\ne\sigma(\hH_1)$ for two arbitrary Hamiltonians $\hH_0$ and $\hH_1$  in one of the classes C1, C2, or C3, they can never be  smoothly connected within the same class.

In order to connect such $\hH_0$ and $\hH_1$ within the same class, one must go through a phase transition, either by passing through a gapless model or a model with nonunique ground states, or by experiencing a discontinuous jump in the expectation value $\omega_s(\hA)$ of a certain local operator $\hA$.

Consider, as an example, the AKLT model $\HAKLT=\sum_j\{\bS_j\cdot\bS_{j+1}+(\bS_j\cdot\bS_{j+1})^2/3\}$, which is in C1, C2, and C3.
The model has a unique ground state $\omega_{\rm VBS}(\cdot)$, called the VBS state, and a nonzero energy gap \cite{AKLT2,TasakiBOOK}.
As we see below, it can be shown that $\omega_{\rm VBS}(\hV_\ell)\simeq-1$ for sufficiently large $\ell$, and hence $\sigma(\HAKLT)=-1$.
There are many examples, including the trivial model \newline$\Htr=\sum_j(\Sz)^2$ and the dimerized model $\hH_{\rm dim}=\sum_k\bS_{2k}\cdot\bS_{2k+1}$, which are in C1, C2, and C3, have a unique ground state with a gap, and characterized by the index $\sigma(\hH)=1$.
This observation leads to the following corollary, whose special case is the conclusion about a phase transition in  \rlb{Ht}. 

\para{Corollary 2}%
One must go through a phase transition in order to connect $\HAKLT$ to $\Htr$ or $\hH_{\rm dim}$ (or other Hamiltonians with trivial index) within one of the classes C1, C2, or C3.
Thus the AKLT Hamiltonian is in a nontrivial symmetry protected topological phase.

\midskip\para{Proof of the theorems}%
We start from a variational estimate of the Lieb-Schultz-Mattis type.

\para{Lemma}%
There are positive constants $C$ and $\ell_0$ which depend only on the constants $r$ and $h_0$ (which we fixed in the beginning).
For any Hamiltonian  $\hH$ (in C1, C2, or C3) and its (not necessarily unique) ground state  $\omega(\cdot)$, we have for any $\ell\ge\ell_0$ that
\eq
\omega(\hV_\ell^\dagger[\hH,\hV_\ell])=\omega(\hV_\ell^\dagger\hH\hV_\ell-\hH)\le\frac{C}{\ell}.
\lb{oVHV}
\en
\noindent
{\em Proof}\/:
Following \cite{Koma2000}, we note that $\omega(\hV_\ell\hH\hV_\ell^\dagger-\hH)\ge0$ because $\omega(\cdot)$ is a ground state.
Then
\eqa
\omega(\hV_\ell^\dagger\hH\hV_\ell-\hH)&\le\omega(\hV_\ell^\dagger\hH\hV_\ell+\hV_\ell\hH\hV^\dagger_\ell-2\hH)
\nl
&=\!\!\!\!\sum_{j:|j-\frac{1}{2}|\le\ell+r+\frac{1}{2}}\!\!\!\!
\omega(\hV_\ell^\dagger\hh_j\hV_\ell+\hV_\ell\hh_j\hV^\dagger_\ell-2\hh_j)
\nl
&\le\!\!\!\!\sum_{j:|j-\frac{1}{2}|\le\ell+r+\frac{1}{2}}\!\!\!\!
\Vert\hV_\ell^\dagger\hh_j\hV_\ell+\hV_\ell\hh_j\hV^\dagger_\ell-2\hh_j\Vert.
\lb{LSM}
\ena
Define the local twist operator around $j$ as $\hV_{j,\varepsilon}=\bigotimes_{k:|k-j|\le r}\exp[-i\varepsilon(k-j)\Sz_k]$.
By using the rotation invariance of $h_j$, we find 
$\hV_\ell^\dagger\hh_j\hV_\ell+\hV_\ell\hh_j\hV^\dagger_\ell=
\hV_{j,\varepsilon}^\dagger\hh_j\hV_{j,\varepsilon}+\hV_{j,-\varepsilon}^\dagger\hh_j\hV_{j,-\varepsilon}$ with $\varepsilon=\pi/\ell$.
Note that this is an even function of $\varepsilon$ which equals $2\hh_j$ when $\varepsilon=0$.
Thus by expanding in $\varepsilon$ and using $\Vert\hh_j\Vert\le h_0$, we find $\Vert\hV_{j,\varepsilon}^\dagger\hh_j\hV_{j,\varepsilon}+\hV_{j,-\varepsilon}^\dagger\hh_j\hV_{j,-\varepsilon}-2\hh_j\Vert\le B\varepsilon^2$ for sufficiently small $\varepsilon$ with a constant $B$ which depends only on $r$ and $h_0$.
Note that the local Hamiltonians near $\pm\ell$ satisfies the same bound since they are less modified.
Thus the right-hand side of \rlb{LSM} is bounded by $2(\ell+r+1)B(\pi/\ell)^2$, which is further bounded by $C/\ell$ for sufficiently large $\ell$.~\qedm

We  now prove the theorems.
Note that, since we are always dealing with a unique ground state  of a Hamiltonian in C1, C2, or C3, the expectation value of $\hV_\ell$ is real.

To prove Theorem~1, we treat $\ell$ as a continuous variable, and assume that the expectation value $\omega(\hV_\ell)$ changes its sign for $\ell$ such that $\ell>\max\{\ell_0,C/\DE\}$.
Note that $\hV_\ell$ is continuous in $\ell$ as an operator because $\exp[i\,2\pi\Sz_j]=1$.
Since its expectation value  $\omega(\hV_\ell)$ is also continuous in $\ell$, there must be $\ell_1$ with $\ell_1>\max\{\ell_0,C/\DE\}$ such that $\omega(\hV_{\ell_1})=0$.
But this, with the variational estimate \rlb{oVHV}, contradicts the assumption that the gap is $\DE$.
Recall \rlb{gap} and note that $C/\ell_1<\DE$.

The proof of Theorem~2 is similar.
Suppose that $\hH_0$ and $\hH_1$ are smoothly connected with the minimum gap $\DE_{\rm min}>0$.
We then choose $\ell$ so that $\ell\ge\ell_0$ and $C/\ell<\DE_{\rm min}$.
We find from the assumed continuity of $\omega_s(\hV_\ell)$ that there is $s\in(0,1)$ such that $\omega_s(\hV_\ell)=0$.
Again this contradicts the assumption that the minimum gap is $\DE_{\rm min}$.

\midskip
It remains to verify two minor points.
Let us show the reality of $\omega(\hV_\ell)$ for C3.
This is not as straightforward as the other two classes.
We work on a finite chain $\{-L,\ldots,L+1\}$.
Let $\ket{\Phi^{\rm gs}}=\sum_{\bss}\varphi(\bss)\ket{\bss}$ be a ground state, where $\varphi(\bss)\in\bbC$ is a coefficient, and $\ket{\bss}$ is the standard $\hat{S}^{\rm z}$-basis states corresponding to a spin configuration $\bss=(\sigma_{-L},\ldots,\sigma_{L+1})$ with $\sigma_j=0,\pm1$.
The time-reversal invariance of the Hamiltonian implies that the time-reversal of $\ket{\Phi^{\rm gs}}$ given by $\ket{\Psi^{\rm gs}}=\sum_{\bss}\{\prod_{j=-L}^{L+1}(-1)^{1+\sigma_j}\}\{\varphi(-\bss)\}^*\ket{\bss}$ also converges to the same infinite volume ground state $\omega(\cdot)$.
Then it is easily confirmed that $\omega(\hV_\ell)\in\bbR$ by comparing the expressions for $\bra{\Phi^{\rm gs}}\hV_\ell\ket{\Phi^{\rm gs}}$ and $\bra{\Psi^{\rm gs}}\hV_\ell\ket{\Psi^{\rm gs}}$.

Let us explain one of many methods to evaluate the expectation value $\omega_{\rm VBS}(\hV_\ell)$ for the VBS state, the exact ground state of $\HAKLT$.
See also \cite{NakamuraTodo2002,TasakiNext}.
We assume for simplicity that $\ell$ is an integer.
It is known that, in the VBS state, configurations of the z-component of spins on a finite interval $\{-\ell,\ldots,\ell+1\}$ is exactly obtained as follows \cite{AKLT2,TasakiBOOK}.
For each site one assigns spin 0 with probability 1/3, or leave it unspecified with probability 2/3.
This is done independently for all sites in the interval.
Then to the unspecified sites, one assigns a completely alternating sequence $+$$-$$+$$-$$+$$-$$\cdots$ or $-$$+$$-$$+$$-$$+$$\cdots$, each with probability 1/2.
In this way one gets spin configurations without conventional order but with hidden antiferromagnetic order.
For a given configuration $(\sigma_{-\ell},\ldots,\sigma_{\ell+1})$ with $\sigma_j=0,\pm$, we let $j_1,\ldots,j_N$ be those sites with $\sigma_j=\pm$, ordered as $j_k<j_{k+1}$.
Then one finds by inspection that
\eq
\sum_{j=-\ell}^{\ell+1}(j+\ell)\sigma_j=-\sigma_{j_1}\sum_{k=1}^{[N/2]}(j_{2k}-j_{2k-1})+\chi_{\rm odd}(j_N+\ell)\sigma_{j_N},
\en
where $\chi_{\rm odd}=1$ or 0 if $N$ is odd or even, respectively.
Note that $\sum_{k=1}^{[N/2]}(j_{2k}-j_{2k-1})$ may be interpreted as the polarization by identifying $\pm$ spins with $\pm$ charges.
Since sites $j_1,\ldots,j_N$ are chosen randomly, we see for large $\ell$ that $j_N\simeq\ell$ , and  $\sum_{k=1}^{[N/2]}\langle j_{2k}-j_{2k-1}\rangle\simeq\ell$, where $\langle\cdots\rangle$ denotes the average with respect to the probability described above.
Then the law of large number implies that
\eq
\exp\Bigl[-i\,2\pi \sum_{j=-\ell}^{\ell+1}\frac{j+\ell}{2\ell+1}\,\sigma_j\Bigr]\to-1,
\en
as $\ell\up\infty$ with probability 1.

\midskip\para{Discussion}%
In $S=1$ quantum spin chains, we have shown that the expectation value of the Affleck-Lieb twist operator, a local version of the Nakamura-Todo order parameter \cite{NakamuraTodo2002},  defines a $\bbZ_2$ index for a  unique gapped ground state.
The index enables us to prove that the AKLT model cannot be smoothly connected to a trivial model with index 1 within the class  C1, C2, or C3.
As far as we know, this is the first rigorous demonstration of the existence of a symmetry protected topological phase in an interacting quantum many-body system.
Our proof also shows why and how a gapless mode emerges at the transition point.

The reader may have noticed that the three classes of Hamiltonians, C1, C2, and C3, correspond to the three types, S1, S2, and S3, of symmetry necessary to protect the Haldane phase. See Introduction.
Note however that we are requiring the rotational symmetry about the z-axis.
From the point of view of symmetry protected topological phase, the additional rotational symmetry is not only unnecessary but may lead to more complicated phase structures \cite{Kapustin}.
The requirement of the rotational symmetry is certainly not desirable.
But, for the moment, the symmetry seems to be indispensable for our proof, which makes use of the Lieb-Schultz-Mattis type argument.
Recent (not yet rigorous) Lieb-Schultz-Mattis type statements without a continuous symmetry \cite{WPVZ} may contain a hint for removing the assumption.
We note however that our index, or, equivalently, the Nakamura-Todo order parameter likely fails to distinguish the Haldane phase when only the symmetry S2 or S3 is present.

Although we have concentrated on $S=1$ chains for simplicity, all the general results in the present paper readily extend to spin chains with general $S$ \cite{TasakiNext}.
A nontrivial point is whether we can identify ground states with nontrivial index $-1$.
Tractable examples include various VBS type states \cite{AKLT2}, including the VBS state for odd $S$, intermediate $D$ state  \cite{Oshikawa92} for $S=2$, and the partially magnetized VBS state \cite{Oshikawa92,OYA} for $S=2$.
The extension to lattice electron systems show that the AKLT model cannot be smoothly connected to a trivial band insulator when proper symmetry is assumed \cite{TasakiNext}.

\bigskip
{\small
I wish to thank Tohru Koma and Ken Shiozaki for discussions which inspired the present work, and for indispensable comments.
I also thank
Yohei Fuji,
Yasuhiro Hatsugai,
Hosho Katsura,
Bruno Nachtergale,
Masaaki Nakamura,
Masaki Oshikawa,
and 
Haruki Watanabe
for valuable discussions and comments.
The present work was supported by JSPS Grants-in-Aid for Scientific Research no.~16H02211.
}


\end{document}